\documentclass[useAMS,usenatbib]{mn2e}
 
\usepackage{amsmath}
\usepackage{float}
\usepackage{graphicx}
\usepackage{txfonts}
\usepackage{indentfirst}
\usepackage{color}
\usepackage{ulem}
\usepackage{hyperref}
 
\newcommand{\eq}[1]{\begin{equation}  #1 \end{equation}}
\newcommand{\eqs}[1]{\begin{equation} \begin{split} #1 \end{split} \end{equation}}

\newcommand{\br}[1]{\left( #1 \right)}

\newcommand{\bb}[1]{\left[ #1 \right]}

\newcommand{\dd}{{\rm d}}

\newcommand{\rund}[1]{\left(#1\right)}

\def\araa{ARA\&A}
\def\apj{ApJ}

\def\aap{A\&A}

\def\mnras{MNRAS}
\def\nat{Nature}
\def\na{Nature}
\def\aplett{Astrophys.~Lett.}

\def\aapr{A\&A~Rev.}

\title[Analytical model for non-thermal pressure in galaxy clusters]{Analytical
model for non-thermal pressure in galaxy clusters}

\author[Xun Shi and Eiichiro Komatsu]{Xun Shi$^{1}$\thanks{E-mail:
xun@mpa-garching.mpg.de} and Eiichiro Komatsu$^{1,2}\thanks{E-mail:
komatsu@mpa-garching.mpg.de}$\\
$^{1}$Max-Planck-Institut f\"ur Astrophysik,
Karl-Schwarzschild-Stra{\ss}e 1, 85740 Garching bei M\"unchen, Germany\\
$^{2}$Kavli Institute for the Physics and
Mathematics of the Universe, Todai Institutes for Advanced Study, the
University of Tokyo, Kashiwa, Japan 277-8583\\ (Kavli IPMU, WPI)}

\begin{document}

\pagerange{\pageref{firstpage}--\pageref{lastpage} \pubyear{2014}}

\maketitle

\label{firstpage}

\begin{abstract}
Non-thermal pressure in the intracluster gas has been found
ubiquitously in numerical simulations, and observed indirectly. 
In this paper we develop  an
analytical model for intracluster non-thermal pressure in the
virial region of relaxed clusters.
We write down and solve a first-order differential equation
describing the evolution of non-thermal velocity dispersion.  
This equation is based on insights gained from observations,
numerical simulations, and theory of turbulence.  
The non-thermal energy 
is sourced, in a self-similar fashion, by the mass growth
of clusters via mergers and accretion, and dissipates with a time-scale
determined by the turnover time of the largest turbulence eddies. 
Our model predicts a radial profile of non-thermal pressure
for relaxed clusters. 
The non-thermal fraction increases with radius, redshift, and cluster
mass, in agreement with 
numerical simulations. The radial dependence is due to a
rapid increase of the dissipation time-scale with radii, and the mass
and redshift dependence comes from the mass growth history.
Combing our model for the non-thermal fraction with the Komatsu-Seljak 
model for the total pressure, we obtain
thermal pressure profiles, and compute the hydrostatic mass
bias. We find typically 10\% bias for the hydrostatic mass enclosed
within $r_{500}$. 

\end{abstract}


\section[]{Introduction}
During hierarchical assembly of galaxy clusters through accretion and
mergers, gravitational energy is transferred to random motion of the
intracluster matter.  
Most of the random motion of baryonic matter 
is thermalized, leading to X-ray emitting temperatures as well as high
thermal pressure support which sustains the intracluster gas against
a gravitational collapse.  

In addition to the thermal motion, there can be other
sources of pressure support in 
the intracluster gas, including bulk motion, turbulence, cosmic rays,
and magnetic fields.
The non-thermal pressure support is difficult to detect
directly using 
X-ray or Sunyaev-Zel'dovich (SZ) observations due to the small electron
velocities it is associated with. 
However, if not accounted for, it causes a 
bias in the cluster mass estimated from X-ray or SZ
observations assuming hydrostatic equilibrium between gravity and
thermal, rather than total, pressure.
 
While observational evidence of intracluster non-thermal pressure
exists \citep[see e.g.,][and references within]{allen98, mah08, richard10, zhang10,
shang12, vdl14}, most of the current knowledge on the level of non-thermal
pressure support comes from numerical simulations \citep[e.g.,][]{dolag05,vazza06,vazza09,iapichino08,maier09,iapichino11,
shaw10, bat12}. The simulations show a common trend of  non-thermal
fraction increasing towards larger radii in the cluster
outskirts, and becoming  
comparable to the thermal
pressure support at the virial radius. In terms of the cluster mass
estimation,
a hydrostatic mass bias is usually found to be on the order of $5-20\%$  
in simulations
\citep[e.g.,][]{rasia06,rasia12,nagai07b,piff08,men10,conte11,nelson12,krause12}.

In this paper, we present an analytical model for  
the non-thermal pressure support, which combines the insights gained from observations and numerical simulations.

The rest of the paper is organized as follows.
In Sect.\;\ref{sec:existing}, we review the existing knowledge of
non-thermal energy density in the
intracluster gas, and draw reasonable postulations from them. 
In Sect.\;\ref{sec:main}, we  
present a model for the fraction of non-thermal
pressure relative to the total in
the intracluster gas. 
In Sect.\;\ref{sec:Ptot}, we combine our model for the
non-thermal fraction with a model for the total pressure to
derive non-thermal and thermal pressure profiles.  
In Sect.\;\ref{sec:res}, we compare the model predictions with 
simulations and observations. 
In Sect.\;\ref{sec:bias} we calculate the hydrostatic mass bias. 
We discuss other sources of non-thermal pressure support
in Sect.\;\ref{sec:dis}, discuss future research directions in
Sect.\;\ref{sec:test}, and conclude in Sect.\;\ref{sec:con}.

We use the following cosmological parameters of  
a flat $\Lambda$CDM cosmology: 
matter content $\Omega_{\rm
m0}=0.28$, dark energy content $\Omega_{\Lambda}=0.72$, Hubble parameter
$h_0=0.7$, slope of the initial power spectrum $n_s=0.96$, and the
normalization of the matter power spectrum $\sigma_8=0.8$.

\section{Injection and dissipation of intracluster turbulence}
\label{sec:existing}

We decompose the total pressure into  
a thermal (`th') and a non-thermal (`nth') part,   
$P_{\rm tot} \equiv P_{\rm th} + P_{\rm nth}$. 
We focus our study on what is usually considered as  
the major source of
$P_{\rm nth}$: the non-thermal random motion in the intracluster gas, and refer to it as `turbulence' without distinguishing it
from isotropic bulk motions. Other possible sources of non-thermal
pressure, e.g., magnetic fields and cosmic rays, are neglected for the
moment. We shall discuss them later in
Sect.\;\ref{sec:dis}. 

We write the non-thermal pressure 
as ${P_{\rm nth} = \rho_{\rm gas} \sigma^2_{\rm nth}}$, where $\rho_{\rm
gas}$ is the mass density of intracluster gas and  $\sigma_{\rm nth}$ is the
one-dimensional velocity dispersion of the non-thermal random motion.
Similarly, we write the thermal pressure 
as ${P_{\rm th} = \rho_{\rm gas} \sigma^2_{\rm th}}$, where 
 $\sigma_{\rm th}$ is the one-dimensional velocity dispersion of the
 thermal motion. 
The total velocity dispersion  is given by
$\sigma^2_{\rm tot}\equiv \sigma^2_{\rm th}  + \sigma^2_{\rm nth}$.
We then define the `non-thermal fraction,' $f_{\rm nth}$, as
\eq{ f_{\rm nth} \equiv \frac{P_{\rm nth}}{P_{\rm th} + P_{\rm nth}} =
\frac{\sigma^2_{\rm nth}}{\sigma^2_{\rm tot}} \,. }
As $\sigma^2_{\rm th}$ and $\sigma^2_{\rm nth}$ can
be regarded as  
thermal and turbulence energies per unit mass per degree of freedom,
respectively, the non-thermal fraction reflects the
evolution of turbulence and thermal energies in the intracluster medium.

Note that turbulent motions exist in the intracluster medium only
as a transient phenomenon. Given a sufficiently long time, they will all dissipate
into heat, and the intracluster gas will be in perfect hydrostatic equilibrium.
Therefore, how much turbulence there is in a cluster depends predominantly on
its dynamical state, and the amplitude of intracluster turbulence
reflects the past and undergoing growth of the cluster via accretion and merger
events.

\subsection{Turbulence injection}
During the continuous growth of a galaxy cluster in mass,
gravitational energy is converted into kinetic energy, and then
further into turbulence and thermal energy. At each position in the cluster, we
write the partition of turbulence and thermal energy increases as  
\eq{
\label{eq:etadefi}
\Delta \sigma^2_{\rm nth} = \eta \;\Delta \sigma^2_{\rm tot} = \eta \;\br{
\Delta \sigma^2_{\rm th} + \Delta \sigma^2_{\rm nth}} \,,
}
i.e., during cluster growth, the increase in 
turbulence energy is $\eta/(1-\eta)$  
times the increase in thermal energy for a local unit of intracluster gas.

Physically, kinetic energy can be converted into turbulence and thermal energy
through different processes, e.g., merger and accretion shocks and
wakes, cluster core sloshing and galaxy wakes, with the help of plasma
instabilities. 
Gravitational energy can also be adiabatically converted into thermal
and turbulence energy via 
adiabatic contraction. Supported by hydrodynamic simulations
\citep[e.g.,][]{ryu03,pfr06,ski08}, we consider the low Mach-number intracluster
shocks created at merger events or merger-induced complex flows called `internal
shocks' to be the major machinery of energy conversion.
Consequently, we shall call the parameter $\eta$ in
equation~(\ref{eq:etadefi}) the `turbulence injection efficiency' of
the internal shocks. This parameter should be determined primarily by the Mach
number of the internal shocks. 

We shall assume that $\eta$ is a constant. While it is
mainly for simplicity, this assumption is supported by the fact that 
the Mach numbers of the merger induced shocks are
expected to be nearly universal from simple physical arguments \citep[e.g.,][and
references therein]{sar02}. In simulations, the redshift and radius dependencies
of the internal shock Mach number are indeed found to be weak within the virial
region of the cluster (see, e.g., Fig.\;6 of \citet{ryu03} for the redshift
dependence, and Fig.\;9 of \citet{vazza09} for the radius dependence; see also
\citet{vazza11} for a comparison of different simulation methods, although at
lower resolutions). We shall discuss the consequence of additional dependence
of $\eta$ in Sect.\;\ref{sec:dis}.

The energy injected into the intracluster gas during the cluster mass
growth gets distributed over different radii as follows. To the first order,
the intracluster pressure profile, and thus the
energy density profile of the intracluster gas, is self-similar
\citep{nagai07,arnaud10}.
To maintain this self-similarity during the mass growth of
a cluster, the energy 
density must increase in a multiplicative fashion, i.e., the energy gained by a
gas particle in the cluster is proportional to the existing
energy of that particle, $\Delta \sigma^2_{\rm tot} \propto \sigma^2_{\rm tot}$.
It has been shown numerically that this near
self-similarity holds 
even after violent major-merger events 
\citep{mcc08}. This can again be understood
physically by the fact that the
intracluster gas is processed by shocks whose Mach numbers do not vary
very much as the shocks travel across the whole cluster.

In previous analytical studies, different turbulence energy injection
scenarios have been considered. 
\citet{cas05} consider turbulence injection at single merger events, and 
assume that a fraction of the $PdV$ work done by the in-falling sub-cluster is
converted into intracluster turbulence energy. The volume swept by the
sub-cluster in which turbulence is injected is estimated from the virial
radius of the main cluster and the radius of the sub-cluster after ram-pressure
stripping, following \citet{fuj03}.
In their model, the turbulence injected roughly scales with the thermal
energy of the cluster, which is in accordance with the $\Delta
\sigma^2_{\rm nth}  \propto \Delta 
\sigma^2_{\rm tot} \propto \sigma^2_{\rm tot}$ relation in our considerations,
although they do not consider the radial distribution of the energy.

\citet{cav11} also consider the non-thermal pressure support in the
intracluster medium contributed by turbulence.
In their work, turbulence is driven by the inflows of
intergalactic gas across the virial accretion shocks. They assume an exponential
decay of the radial profile of the turbulence-to-thermal fraction. The
correlation length and the amplitude of the profile need to be obtained from
fitting to observations. Since their turbulence driving mechanism is
 very different from what we consider, their results on turbulence pressure
 differ from ours, which we will discuss in Sect.\;\ref{sec:nthfrac}.

In addition to the structure formation process, magnetothermal instability may drive turbulence in the presence of intracluster magnetic fields
and a long-standing negative temperature gradient \citet{par12, mcc13}. The 
turbulence driven by magnetothermal instability is found to be generally
additive to the turbulence produced by structure formation. We shall not
consider this potential source of non-thermal pressure in
this paper. 

\subsection{Turbulence dissipation}

Once injected, turbulence in the intracluster gas starts to dissipate. This
process can be described by the standard theory of dissipation of turbulence
\citep[see, e.g.,][]{landau59}.  
Any kinetic energy on the viscous scale (the so-called
Kolmogorov microscale) dissipates rapidly into heat. The overall
time-scale of dissipation of turbulent kinetic energy, $t_{\rm d}$, is
controlled by the efficiency of energy transfer from
the largest scales, where most of the turbulence energy is stored,
to smaller scales. This time-scale is 
linearly related to the eddy turn-over time of the largest eddies. 
Consider the largest eddy at a radius $r$ from the cluster center to have a
size proportional to $r$, and a typical peculiar
velocity on the order of the local orbital velocity given by $v(r) = r\Omega(r) =
\sqrt{G M(<r)/r}$, with $\Omega$  
being the corresponding angular velocity, and $M(<r)$ the cluster mass
within a radius $r$. The dissipation time-scale, $t_d$, is then
proportional to the local dynamical time,  
i.e., $t_d\propto t_{\rm dyn} \equiv 2\pi/\Omega$. We write
\eq{
\label{eq:td}
t_{\rm d} \equiv \frac{\beta}{2}\; t_{\rm dyn} = \pi\beta\; \Omega^{-1} = \pi
\beta \sqrt{\frac{r^3}{G M(<r)}} \,,
}   
where $\beta$ is a coefficient determined by microphysics. This definition of
$t_{\rm d}$ allows the preferred value of $\beta$ to be around unity, see
Sect.\;\ref{sec:res}.

To compute equation~(\ref{eq:td}), we use $M(<r)$ computed
from the NFW density profile \citep{nfw96,nfw97},  
\eq{
\label{eq:mmnfw}
M(<r,M_{\rm vir},t) =  M_{\rm vir} \frac{m(c r/r_{\rm vir})}{m(c)}  \,,
}
where
\eq{
\label{eq:smallm}
m(x) \equiv \ln\br{1+x}-\frac{x}{1+x} \,,
} 
and
$M_{\rm vir}$ and $r_{\rm vir}(M_{\rm vir},t)$ are the virial mass and
radius, respectively, and $c(M_{\rm vir},z(t))$ is the concentration
parameter, for which we use the fitting formula developed by \citet{duffy08}.

The relation, $t_{\rm d} \propto t_{\rm dyn}$, derives from our assumption
on the largest eddy size at each radius. Large coherent motions 
can be easily broken into eddies with sizes proportional to $r$ owing to the
geometry and density structure of the cluster. Whether the eddies at cluster
outskirts could reach such large sizes depends on the turbulence injection
mechanism. In hydrodynamical simulations, large-scale flows and low-curvature
shocks are observed at cluster outskirts \citep[e.g.,][]{vazza10}, enabling the
formation of large size turbulence eddies.

If intracluster turbulence is coupled with magnetic fields,
turbulence can give rise to magnetohydrodynamical waves which transfer
energy from magnetic fluctuations to relativistic particles on small
scales. This process provides an additional channel for energy
dissipation at high $k$ modes \cite[e.g.][]{bru11}. However, this does
not affect the evolution of turbulence pressure, as the turbulence
energy is contained predominantly in the largest scales which has the
longest timescale. In other words, the turbulence dissipation time-scale
is insensitive to the small scale physics, including the resistivity and
viscosity of the gas.

\section{Evolution of the intracluster non-thermal pressure fraction}
\label{sec:main}
\subsection{The model}
\label{sec:model}

The dynamical time-scales in a galaxy cluster increase  
significantly from the inner region to the outskirts  
due to a steep slope of the density profile.  
It then follows from equation (\ref{eq:td}) that the turbulence 
dissipation time in the inner region of a galaxy cluster is much shorter than
that in the outskirts. As a consequence,  
the magnitude of the non-thermal fraction at each radius evolves
roughly independently with different time-scales.  
The scalings $\Delta \sigma^2_{\rm tot} \propto \sigma^2_{\rm tot}$ and $\Delta
\sigma^2_{\rm nth} = \eta \;\Delta \sigma^2_{\rm tot}$  
suggest that the injection of turbulence energy can also be treated
independently at each radius.
These considerations motivate our focusing  
on the evolution of turbulence and thermal
energies at a single Eulerian radius, $r$.

The turbulence energy dissipates  with a typical time-scale $t_{\rm d}$, and is sourced by the mass growth of
a cluster in a way that $\Delta \sigma^2_{\rm nth} = \eta \;\Delta
\sigma^2_{\rm tot}$.
We thus write the following  
first-order differential equation to describe the evolution of
$\sigma^2_{\rm nth}$ at a given radius $r$,  
\eq{
\label{eq:sigkin}
\frac{\dd \sigma^2_{\rm nth}}{\dd t} = -\frac{\sigma^2_{\rm nth}}{t_{\rm d}}
+ \eta\;\frac{\dd \sigma^2_{\rm tot}}{\dd t}  \,.
}
 In general, the 
terms $\sigma^2_{\rm tot}$, $t_{\rm d}$, and hence  
$\sigma^2_{\rm nth}$,   
are all functions of radius, mass, and redshift of a cluster.

The total one-dimensional velocity dispersion, $\sigma_{\rm tot}$, 
is determined by 
the depth of cluster's gravitational potential.
We present a method to compute $\sigma^2_{\rm tot}$  in
Sect.\;\ref{sec:Ptot}, and regard it as a given quantity here.
Equation (\ref{eq:sigkin}) then forms an initial value
problem. Once the initial condition, $\sigma^2_{\rm nth}(r,M_{\rm i},t_{\rm i})$,  at
the initial time, $t_{\rm 
i}$, is chosen, we can use equation (\ref{eq:sigkin}) to
solve for the evolution of
$\sigma^2_{\rm nth}(r,M_{\rm obs},t_{\rm obs})$ for any later times,
$t_{\rm obs}$, given the mass growth history of the cluster,
$M(t)$. 

The parameters in our model, $\beta$ and $\eta$, have physical origins
and therefore are not free parameters. From the current knowledge of
intracluster gas, however, their values are not yet known precisely.
The turbulence injection efficiency, $\eta$, ranges between $0$ and $1$ by
definition. A high value of $\eta > 0.6$ is expected from the low Mach number
(Mach number $< 3$) of the internal shocks \citep{ryu03,kang07}. 
The parameter $\beta \sim 1$ is expected from theory of turbulence, with rather
large uncertainty.
We will discuss the range of their values in more detail in Sect.\;\ref{sec:res}.

\subsection{Analytical solutions}
\label{sec:analytical}
Three time-scales determine our problem: 
the turbulence dissipation time-scale, $t_{\rm d}$; the time elapsed
between 
the initial time and the time of observation, $t_{\rm obs} - t_{\rm i}$, which in some
situations characterizes the age of the cluster; and a time-scale characterizing
the mass growth rate of the cluster defined by
\eq{
\label{eq:tgrowth}
t_{\rm growth} \equiv \sigma^2_{\rm tot} \br{\frac{\dd \sigma^2_{\rm tot}}{\dd
t}}^{-1}\,.} 
Since $t_{\rm growth}$ also determines the rate at which
turbulence energy grows, we 
refer to it as both the `cluster growth time-scale' and the `turbulence growth
time-scale'. 
The ratios of these three time-scales determine, to a large extent, the value of the non-thermal fraction, $f_{\rm nth}$.

Using $t_{\rm growth}$,  
 equation  (\ref{eq:sigkin}) can be re-written as 
\eq{
\label{eq:evorewrite}
\frac{\dd \sigma^2_{\rm nth}/ \dd t}{\dd \sigma^2_{\rm tot}/ \dd t} = \eta -
\frac{\sigma^2_{\rm nth}}{\sigma^2_{\rm tot}} \frac{t_{\rm growth}}{t_{\rm
d}}\,. }
Let us find an attractor solution defined by   
\eq{
\frac{\dd \sigma^2_{\rm nth}/ \dd t}{\dd \sigma^2_{\rm tot}/ \dd t} =
\frac{\sigma^2_{\rm nth}}{\sigma^2_{\rm tot}} \,.
} Inserting this relation into equation
(\ref{eq:evorewrite}), we find the limiting value of the non-thermal fraction,
\eq{
\label{eq:fnthlim}
f_{\rm nth}^{\rm lim} = \eta \frac{t_{\rm d}}{t_{\rm d}+t_{\rm growth}}\,.
}
It is then apparent that the $f_{\rm nth}^{\rm lim}$ is
smaller than or equal to $\eta$. The upper limit is reached when
intracluster turbulence grows much faster than it dissipates ($t_{\rm
growth} \ll t_{\rm d}$), while in the opposite limit ($t_{\rm
growth} \gg t_{\rm d}$) the non-thermal fraction is very small, $f_{\rm
nth} \to \eta \; t_{\rm d}/t_{\rm growth} \ll \eta$. 
When growth and dissipation time-scales  are
roughly equal ($t_{\rm growth} \approx t_{\rm d}$), $f_{\rm
nth}$ approaches a non-negligible fraction of $\eta$.
In general,  $t_{\rm d}$ and $t_{\rm growth}$ (hence
$f_{\rm nth}$) depend on time, radius, and  mass.
 
Whereas
the ratio of $t_{\rm growth}$ and $t_{\rm d}$ sets
the relative importance of turbulence growth and dissipation, a
comparison of
$t_{\rm obs} - t_{\rm i}$ and the smaller of these two
time-scales, min($t_{\rm growth}$, $t_{\rm d}$), determines
how much turbulence energy can be accumulated by  
the time  of observation. 

To gain some intuition as to  
how the value of $t_{\rm obs} - t_{\rm i}$ affects  
$f_{\rm nth}$, let us take  
$t_{\rm d}$ as a constant.  
We then find a formal solution to equation
(\ref{eq:sigkin}) as   
\eqs{
\label{eq:approxsolution}
\sigma^2_{\rm nth}(t_{\rm obs}) 
 = &\;  \eta \int_{t_{\rm i}}^{t_{\rm obs}}
  \frac{ \sigma^2_{\rm tot}(t)}{t_{\rm growth}(t)} \; \exp\rund{-\frac{t_{\rm
  obs}-t}{t_{\rm d}}}\; \dd t \\
  +& \sigma^2_{\rm nth}(t_{\rm i}) \; \exp\rund{-\frac{t_{\rm obs} - t_{\rm
  i}}{t_{\rm d}}}  \,,
  }
where the first term on the r.h.s. describes accumulation of turbulence
energy within the time span of $t_{\rm obs} - t_{\rm i}$, while the second term
describes dissipation of the initial turbulence energy.

When the cluster growth is fast, 
$t_{\rm growth} \ll t_{\rm d}$, the variation of the integrand 
is dominated by that of
$\sigma^2_{\rm tot}$. When there is enough time for $\sigma^2_{\rm
tot}$ to grow, i.e., $t_{\rm obs} - t_{\rm i}$ is more than a few times
$t_{\rm growth}$, the integral is dominated by the
contribution from $t\approx t_{\rm obs}$. We thus have
\eqs{
\int_{t_{\rm i}}^{t_{\rm obs}}
  & \frac{ \sigma^2_{\rm tot}(t)}{t_{\rm growth}(t)} \; \exp\rund{-\frac{t_{\rm
  obs}-t}{t_{\rm d}}}\; \dd t \approx \;  \int_{t_{\rm i}}^{t_{\rm obs}}
  \frac{ \sigma^2_{\rm tot}(t)}{t_{\rm growth}(t)} \; \dd t\\
   = \;& 
  \sigma^2_{\rm tot}(t_{\rm obs}) - \sigma^2_{\rm tot}(t_{\rm i})  \approx
  \sigma^2_{\rm tot}(t_{\rm obs})\,.
  }  
The second term on the r.h.s. of equation~(\ref{eq:approxsolution})
 becomes negligible in a few cluster growth
time-scales, as $\sigma^2_{\rm nth}(t_{\rm i})\ll \sigma^2_{\rm
 tot}(t_{\rm obs})$. 
Then $\sigma^2_{\rm nth}(t_{\rm obs})
\approx \eta\; \sigma^2_{\rm nth}(t_{\rm tot})$, and the non-thermal fraction
approaches $\eta$, in agreement with the limiting value given in
equation~(\ref{eq:fnthlim}).

When the cluster growth is slow, 
$t_{\rm growth} \gg t_{\rm d}$, the variation of the integrand 
is dominated by that of the exponential. 
In this case, when $t_{\rm obs} - t_{\rm i}$ is more than a few
times $t_{\rm d}$, we find  
\eqs{
& \int_{t_{\rm i}}^{t_{\rm obs}}
  \frac{ \sigma^2_{\rm tot}(t)}{t_{\rm growth}(t)} \; \exp\rund{-\frac{t_{\rm
  obs}-t}{t_{\rm d}}}\; \dd t \\
  \approx \;& \frac{ \sigma^2_{\rm tot}(t_{\rm
  obs})}{t_{\rm growth}(t_{\rm obs})}  \int_{t_{\rm i}}^{t_{\rm obs}}  \exp\rund{-\frac{t_{\rm
  obs}-t}{t_{\rm d}}} \; \dd t\\
   = \;& 
 t_{\rm d} \frac{\sigma^2_{\rm tot}(t_{\rm
  obs})}{t_{\rm growth}(t_{\rm
  obs}) } \bb{1 - \exp\rund{-\frac{t_{\rm obs} - t_{\rm i}}{t_{\rm d}}}} \\
    \approx \;& 
 t_{\rm d} \frac{\sigma^2_{\rm tot}(t_{\rm
  obs})}{t_{\rm growth}(t_{\rm
  obs}) }\,.
  }
The second term on the r.h.s. of equation~(\ref{eq:approxsolution}) is
again negligible, as
the initial turbulence energy  
suffers from a large dissipation.  
The non-thermal fraction thus approaches $\eta \; t_{\rm d}/t_{\rm
growth}$, in agreement with the aforementioned limiting value.

\section{Total pressure support}
\label{sec:Ptot}
To calculate the source term in equation~(\ref{eq:sigkin}),
we need a model for the total pressure, $P_{\rm tot}$. 
We may measure $P_{\rm tot}$ as a sum of the thermal and turbulence
pressure directly from hydrodynamical simulations, or infer it assuming
the hydrostatic equilibrium between the total pressure and gravitational
potential gradients. 

Here, we analytically compute $P_{\rm tot}$ from 
\eq{
\label{eq:hydrostatic}
\frac{1}{\rho_{\rm gas}} \frac{\dd P_{\rm tot}}{\dd r} = - \frac{\dd
\Phi}{\dd r}  \,.
}
Solutions to this equation for a given potential
have been obtained previously under a few different assumptions
\citep{makino98,suto98,kom01}. While the previous work solved this
equation for the thermal gas pressure ($P_{\rm tot}\to P_{\rm th}$ on
the l.h.s.), we shall use this equation to solve for the total pressure.

We shall follow the prescription of \cite{kom01} in this paper.
We assume that the intracluster gas obeys a polytropic equation of
state, $P_{\rm tot}  \propto \rho_{\rm gas}^{\Gamma}$. 
While a polytropic  distribution was introduced by
\citet{lea75} as a power-law relation between $P_{\rm th}$ and $\rho_{\rm gas}$,
recent simulations  find that $P_{\rm tot}$ obeys a
polytropic equation of state better than $P_{\rm th}$
\citep{shaw10,bat12b}.
This assumption is valid to within 10\% precision inside the
virial radius \citep{bat12b}.

With this assumption, we can parametrize the radial distribution of the total
pressure and density of the gas as 
\eq{
\label{eq:Pr}
P_{\rm tot} (r, M, t) = P_0(M, t) \; \theta(r, M, t)^{\Gamma(M, t)/\bb{\Gamma(M,
t)-1}}\,, }
\eq{
\label{eq:rhor}
\rho_{\rm gas} (r, M, t) = \rho_0(M, t) \; \theta(r, M, t)^{1/\bb{\Gamma(M,
t)-1}} \,, }
where $P_0\equiv P_{\rm tot}(r=0)$ and $\rho_0\equiv
\rho_{\rm gas}(r=0)$. 
Inserting these 
forms into equation~(\ref{eq:hydrostatic}), we find   
\eqs{
\label{eq:polyfunc}
\theta(r, M, t) &= 1 + \frac{\Gamma(M, t) -1}{\Gamma(M, t)} \frac{\rho_0(M,
t)}{P_0(M, t)} \bb{\Phi(0, M, t) - \Phi(r, M, t)} \,. \\
}
The gravitational potential profile, $\Phi(r)$, is given by 
\eq{
\label{eq:phi}
\Phi(r,M_{\rm vir},t) 
= - \frac{G M_{\rm vir}}{r_{\rm vir}} \frac{c}{\ln\br{1+c}-c/(1+c)} \frac{\ln
\br{1+c r/r_{\rm vir}}}{c r/r_{\rm vir}}  \,. }
The remaining parameters, $\Gamma$ and $\rho_0/P_0$,  are fixed by
requiring the gas  and dark matter density 
profiles to have the same slope around the virial radius. The resulting
$\Gamma$ and 
$\rho_0/P_0$ are well-approximated as functions of the concentration parameter,
as given by equations (25) and (26) in \citet{kom01}.

The Komatsu-Seljak model 
specifies the radial dependence of the
total pressure and density of the intracluster gas. 
This then allows us to calculate the evolution of
the total velocity dispersion squared, $\sigma^2_{\rm tot}(r, M, t) =
P_{\rm tot}(r, M, t)/\rho_{\rm gas}(r, M, t)$, as 
well as its growth rate at a fixed Eulerian radius, 
\eq{
\label{eq:dsigdtexpression}
\frac{\dd \sigma^2_{\rm tot}(r, M, t)}{\dd t} = \frac{\partial \sigma^2_{\rm
tot}(r, M, t)}{\partial t} +  \frac{\partial \sigma^2_{\rm
tot}(r, M, t)}{\partial M} \frac{\dd M}{\dd t} \,,
}
which yield the source term in equation~(\ref{eq:sigkin}).

Note that $\sigma^2_{\rm tot}$ and its growth rate can be obtained
for individual galaxy clusters only when their individual mass growth
history, $M(t)$, 
is known. This is not a problem when studying numerically simulated clusters.

For observed galaxy clusters,  
individual mass growth histories are not known.  
Therefore, to compare to observational results,  
we study an averaged view of galaxy clusters, i.e.,
we consider `representative clusters' whose mass growth histories are described
by the average mass growth histories of all clusters of the same final mass. 
We shall use the average mass growth histories given by \citet{zhao09}.

\section{Results}
\label{sec:res}

\begin{figure}
\centering
\includegraphics[width=0.45\textwidth]{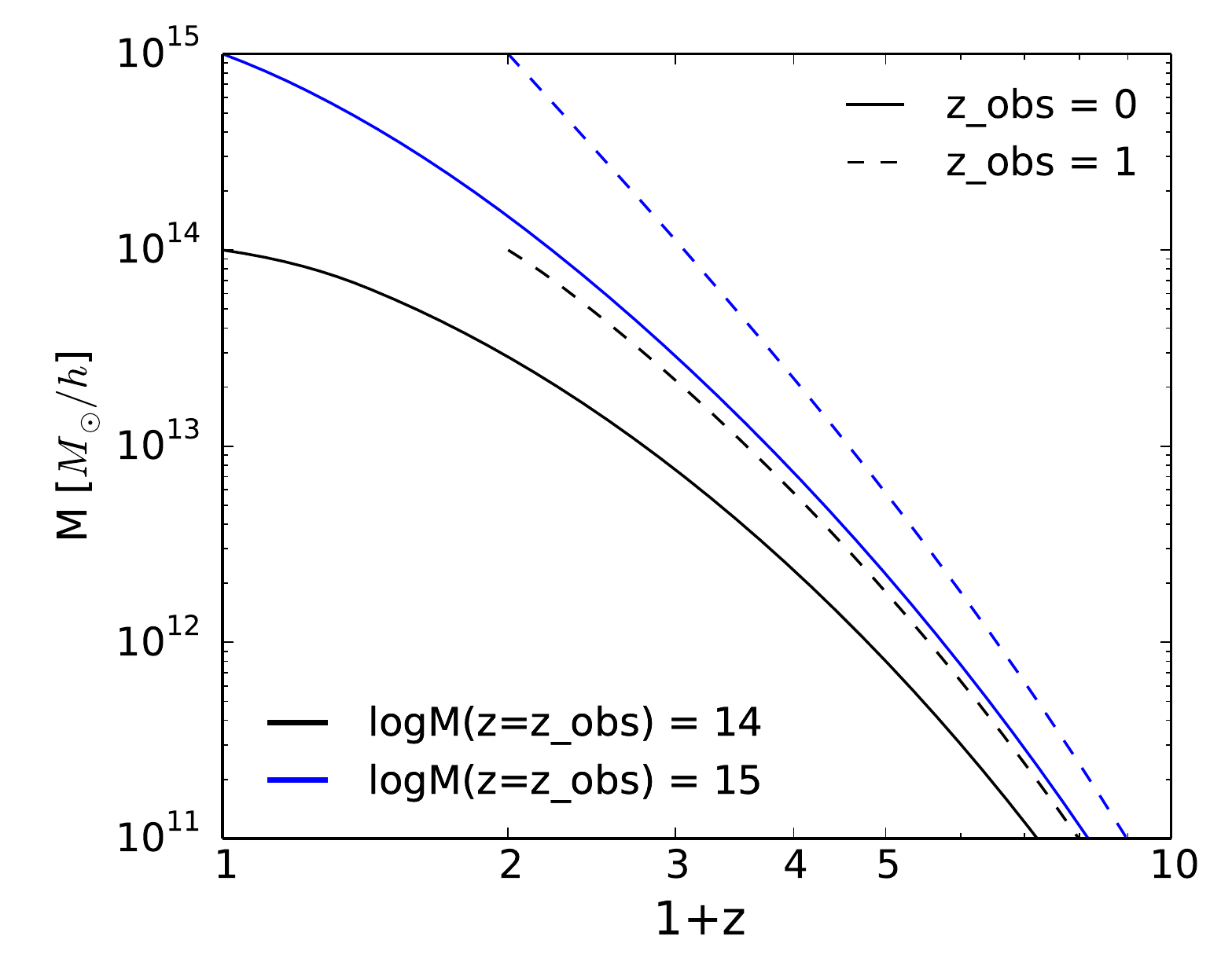}
\caption{Mass growth histories from
\citet{zhao09}. The growth histories for galaxy clusters with the final masses
of $10^{14}~h^{-1}~$M$_{\sun}$ (black) and $10^{15}~h^{-1}~$M$_{\sun}$  (blue)
observed at $z=0$ (solid lines) and $z=1$ (dashed lines) are presented. }
\label{fig:MAH}
\end{figure}

We use equation~(\ref{eq:dsigdtexpression}) with $M(t)$ from
\citet{zhao09} (see Fig.\;\ref{fig:MAH}) and $\sigma^2_{\rm tot}$ from
\citet{kom01} to compute $d\sigma^2_{\rm tot}/dt$, and use it to solve equation~(\ref{eq:sigkin})
for $\sigma_{\rm nth}^2$. 
We choose $\sigma^2_{\rm
nth} = \eta \sigma^2_{\rm tot}$ at $z_{\rm i}=6$ as the initial
condition of equation~(\ref{eq:sigkin}).  We have already argued
that the results are insensitive to a particular choice of
initial conditions. We have confirmed this 
by varying $\sigma^2_{\rm
nth}$ from $0$ to $\sigma^2_{\rm tot}$ at $z_{\rm i}$, and by shifting $z_{\rm
i}$ to higher redshifts.

In the end, our procedure yields non-thermal and thermal
pressure profiles as well as a gas density profile as a function of
radius, mass, and redshift for a given set of  $\beta$ and $\eta$. We
describe our procedure in more detail in Appendix~\ref{app:procedure}.

\subsection{time-scales}

\begin{figure}
\centering
\includegraphics[width=0.45\textwidth]{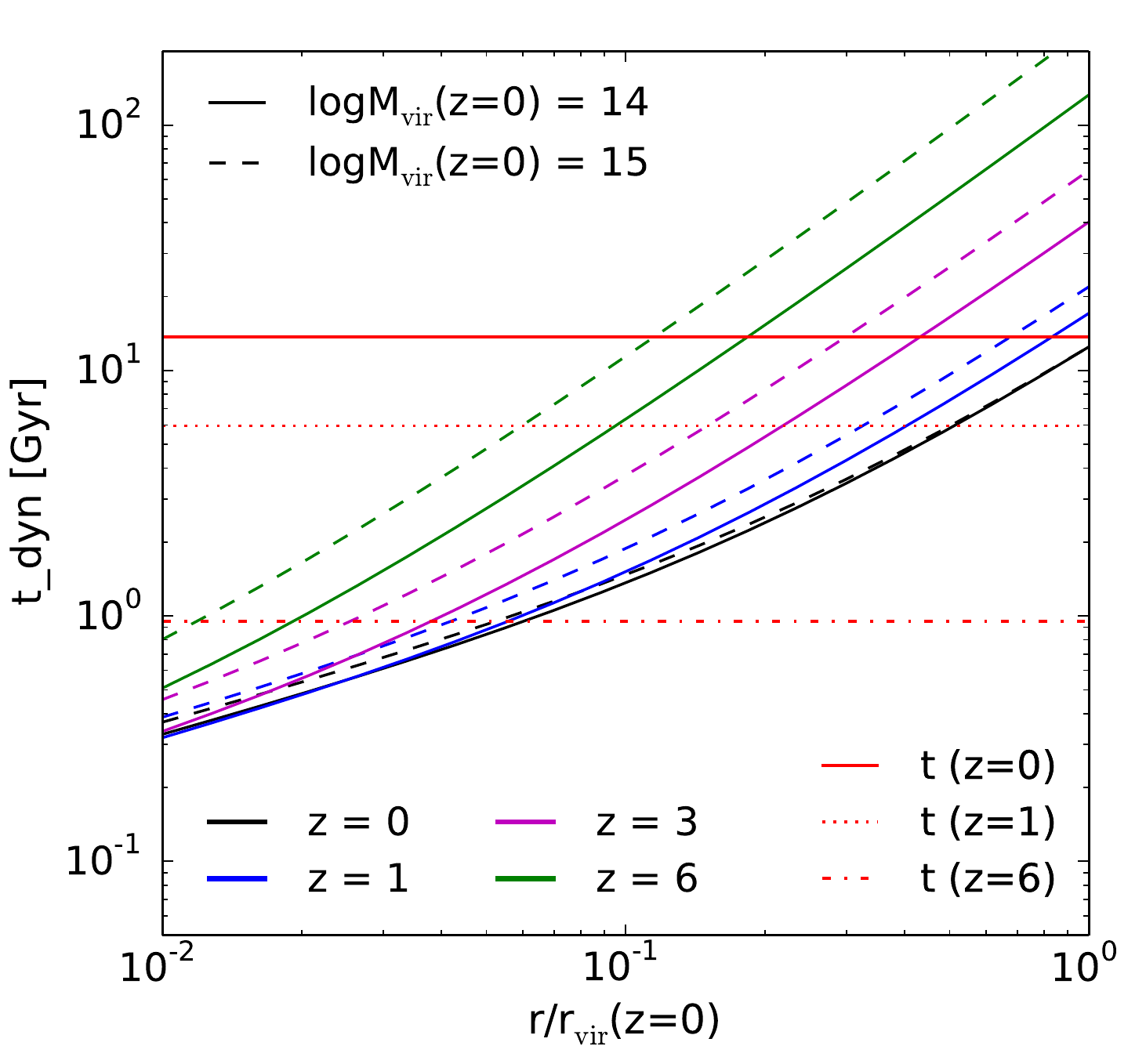}
\caption{Dynamical time,  $t_{\rm dyn}$ (as defined by Eq.\;\ref{eq:td}), of
progenitors of clusters observed at $z=0$ with the final masses of
$10^{14}~h^{-1}~$M$_{\sun}$ (solid lines) and $10^{15}~h^{-1}~$M$_{\sun}$ 
(dashed lines), as they grow in mass with the mean mass growth rate from $z=6$ to $z=0$ (top
to bottom lines; $t_{\rm dyn}$ decreases with $z$).
Shown is $t_{\rm dyn}$ at a fixed Eulerian radius, $r$, divided by the
corresponding virial radius computed  at $z=0$. The horizontal lines show the
proper times (ages) of  the universe at $z=0$, 1, and 6 from the top to the
bottom.}
\label{fig:tdyn}
\end{figure}

\begin{figure}
\centering
\includegraphics[width=0.45\textwidth]{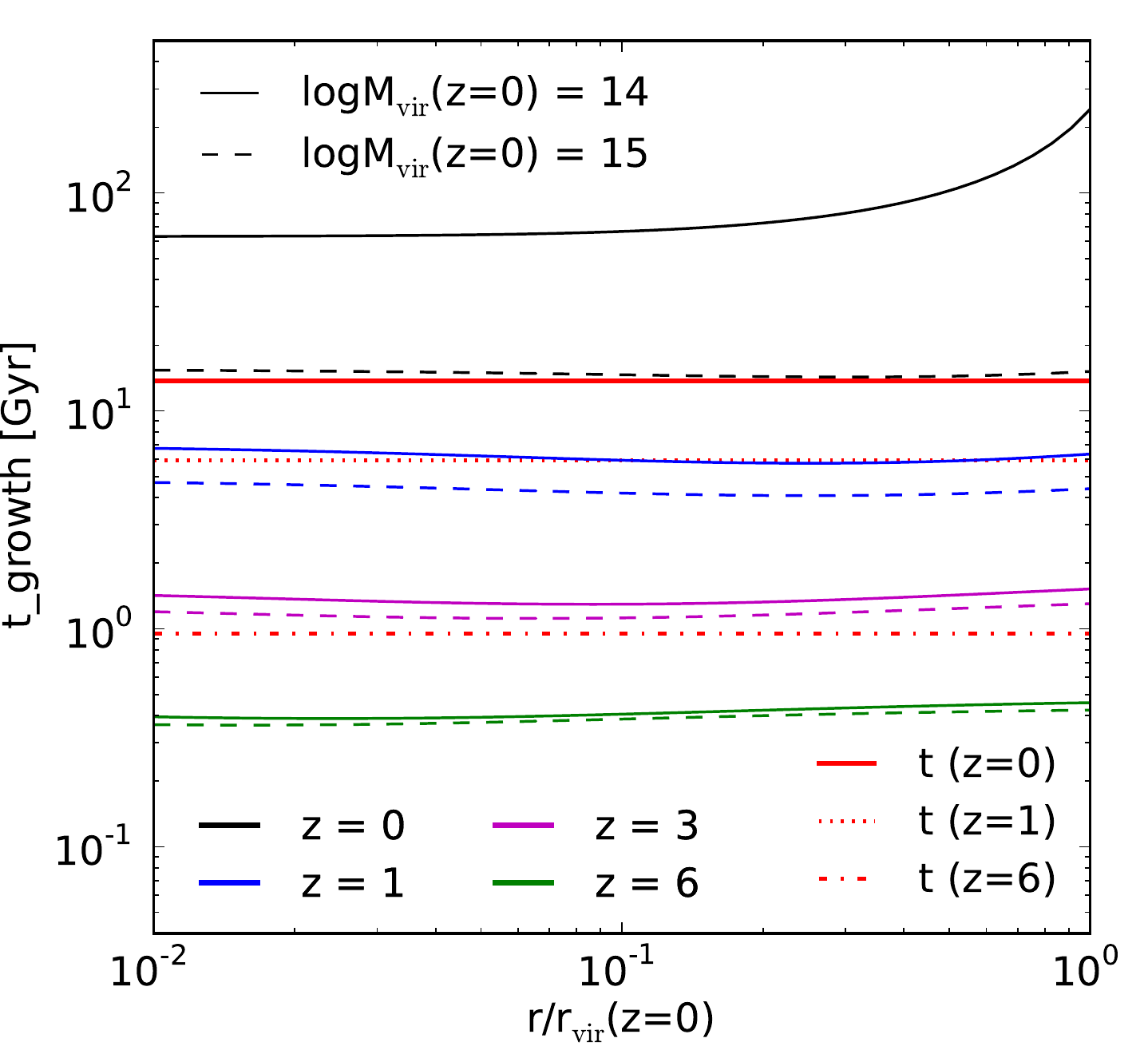}
\caption{Similar to Fig.\;\ref{fig:tdyn}, but for the cluster growth time-scale,
$t_{\rm growth}$, as defined by equation~(\ref{eq:tgrowth}). Note that $t_{\rm
growth}$ increases with $z$.}
\label{fig:tgrowth}
\end{figure}

Before we present the results on non-thermal pressure, let us
study the relevant time-scales of the problem.
In Figs.\;\ref{fig:tdyn} and \ref{fig:tgrowth} we compare
$t_{\rm growth}$ and $t_{\rm dyn} = t_{\rm d}/\beta$ 
to the proper time of the universe, $t$, at a few different redshifts.

The dynamical time-scale given in equation~(\ref{eq:td}),
$t_{\rm dyn}$,  
at one radius is inversely proportional to the mean mass density of all
gravitating matter inside that radius. Due to the steep density profile,
$t_{\rm dyn}$ is much shorter in 
the inner regions than at cluster outskirts. 
At a fixed Eulerian radius, $t_{\rm dyn}$ decreases as
galaxy clusters grow with time for most of the radius and redshift range,
while the typical density of the clusters decreases with time. This
happens because, 
when the cluster is more massive, the same
Eulerian radius approaches the inner region of the cluster. 
The radius at which $t_{\rm dyn} = t$  
increases with time.

The cluster growth time-scale given in
equation~(\ref{eq:tgrowth}), $t_{\rm growth}$,
increases rapidly toward low redshifts. Especially, at
$z=0$, we find $t_{\rm growth}\approx t$ 
for representative clusters with $10^{15}~h^{-1}~$M$_{\sun}$, and 
$t_{\rm growth}\gg t$ for less massive clusters.

Comparing Figs.\;\ref{fig:tdyn} and
\ref{fig:tgrowth}, we find 
$t_{\rm growth}\gtrsim t_{\rm dyn}$
in the inner regions of a cluster
already since $z=6$, and $t_{\rm growth} \gg t_{\rm dyn}$  at low
 redshifts where galaxy clusters are typically observed.
At around the virial radius of clusters at $z=0$, 
we find $t_{\rm growth} \ll t_{\rm dyn}$ at high redshifts
and $t_{\rm growth} \gtrsim t_{\rm dyn}$ at $z=0$. If we choose
$\beta=t_{\rm d}/t_{\rm dyn}\approx 1$, these results imply that
turbulence dissipates 
efficiently in the inner regions of a cluster, while dissipation is less
effective in the cluster outskirts.

As argued in Sect.\;\ref{sec:analytical}, a comparison between $t_{\rm
obs} - t_{\rm i}$ and min$(t_{\rm growth},t_{\rm d})$ determines how
close the non-thermal fraction is to its limiting value $f_{\rm nth}^{\rm
lim}$ given by equation~(\ref{eq:fnthlim}). For the representative
galaxy clusters considered here, we find $(t_{\rm 
obs} - t_{\rm i})/$ min$(t_{\rm growth},t_{\rm d}) > 5$  
at small radii, mostly at low redshifts; thus, we expect  $f_{\rm
nth}^{\rm lim}$  
to be a good approximation of the non-thermal fraction at small radii.

The criterion $(t_{\rm obs} - t_{\rm i})/$ 
min$(t_{\rm growth},t_{\rm d}) > 1$ holds at a much wider redshift and radius
range. At redshifts above $z \approx 1$, it holds for all radii due to the
rapid growth of clusters (hence small $t_{\rm growth}$),  
suggesting that $f_{\rm
nth}^{\rm lim}$  
can also provide a rough estimate of the 
non-thermal fraction at high redshifts for all radii.

\subsection{Non-thermal fraction} 
\label{sec:nthfrac}

As the cluster mass increases and its gravitational potential deepens, the total
velocity dispersion of the intracluster gas grows as well. The turbulence
velocity dispersion, however, depends on a competition of the injection and
dissipation of intracluster turbulence, and does not grow monotonously with
time.

\begin{figure}
\centering
\includegraphics[width=0.45\textwidth]{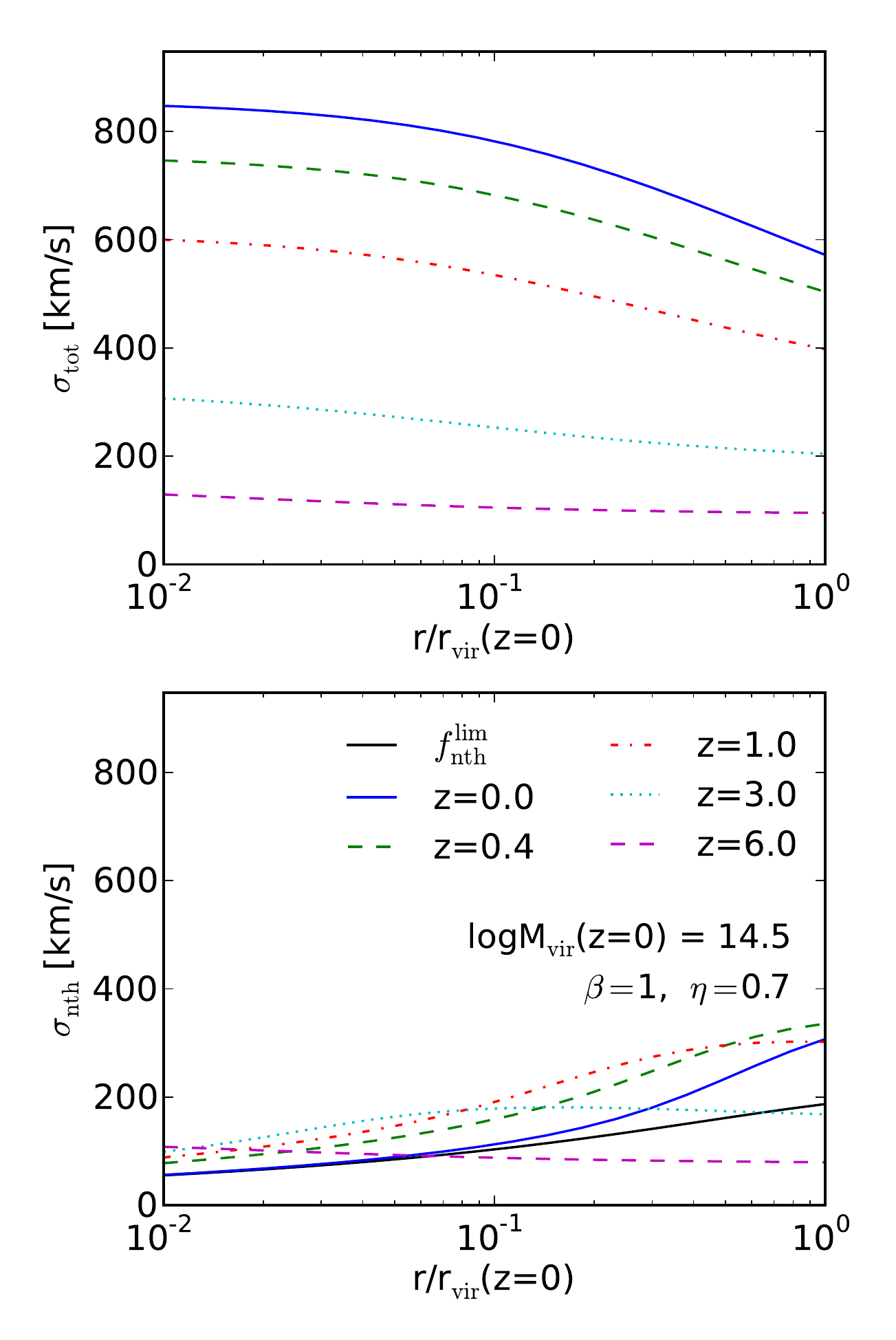}
\caption{Total and non-thermal velocity dispersions, $\sigma_{\rm
tot}$ (top panel) and $\sigma_{\rm nth}$ (bottom panel), in progenitors of
clusters observed at $z=0$ with the final mass of $M_{\rm
vir} = 10^{14.5}~h^{-1}~$M$_{\sun}$.
The velocity dispersions are shown as functions of $z$ and
the Eulerian radius, $r$, divided by the corresponding virial
 radius computed  at $z=0$.
The black solid line in the lower panel
shows the limiting value of the turbulence velocity dispersion, $\sigma_{\rm
nth}^{\rm lim} = \sqrt{f_{\rm nth}^{\rm lim}} \;\sigma_{\rm tot}$, at $z=0$, with
$f_{\rm nth}^{\rm lim}$ given by equation~(\ref{eq:fnthlim}). }
\label{fig:sigmagrowth}
\end{figure}

Fig.\;\ref{fig:sigmagrowth} shows the evolution of the total
(from the Komatsu-Seljak profile) and turbulence
(from equation~\ref{eq:sigkin})
velocity dispersions for 
progenitors of clusters observed at $z=0$
 with the final mass of $10^{\rm 14.5}~h^{-1}~$M$_{\sun}$. The 
coefficient $\beta$ is taken to be unity, and a value of $\eta = 0.7$ is taken for the turbulence
injection efficiency.

The relevant time-scales at $z=0$ are such that
$t_{\rm growth} \gtrsim t_{\rm obs} \gtrsim t_{\rm d}$ at cluster outskirts,
and $t_{\rm obs} \gtrsim t_{\rm growth} \gg t_{\rm d}$ in the inner regions.
At early times ($z\approx 6$), while $t_{\rm d}\lesssim
t_{\rm growth}$
in the inner regions,  
$t_{\rm growth}$ is  
the shortest time-scale at the outskirts.
   
We expect that $\sigma^2_{\rm nth}$ grows
initially with $\sigma^2_{\rm tot}$, but reaches a saturation value and starts
to decrease when $t_{\rm growth}/t_{\rm d} \ge \eta/f_{\rm nth}$ (see equation
\ref{eq:evorewrite}). The above time-scale arguments suggest that this
transition happens earlier in the inner regions than at cluster outskirts. This
behaviour is indeed seen in Fig.\;\ref{fig:sigmagrowth}.

Comparing the limiting value of the non-thermal velocity dispersion (black
solid line) and its actual value (blue solid line) at $z=0$, we find  
that the limiting value is reached in the inner regions, but not at the
outskirts. This is because  
$t_{\rm obs}-t_{\rm i} \gg \rm{min}(t_{\rm growth},
t_{\rm d})$ holds in the inner regions but not at the outskirts.  
 
\begin{figure*}
\centering
\includegraphics[width=0.9\textwidth]{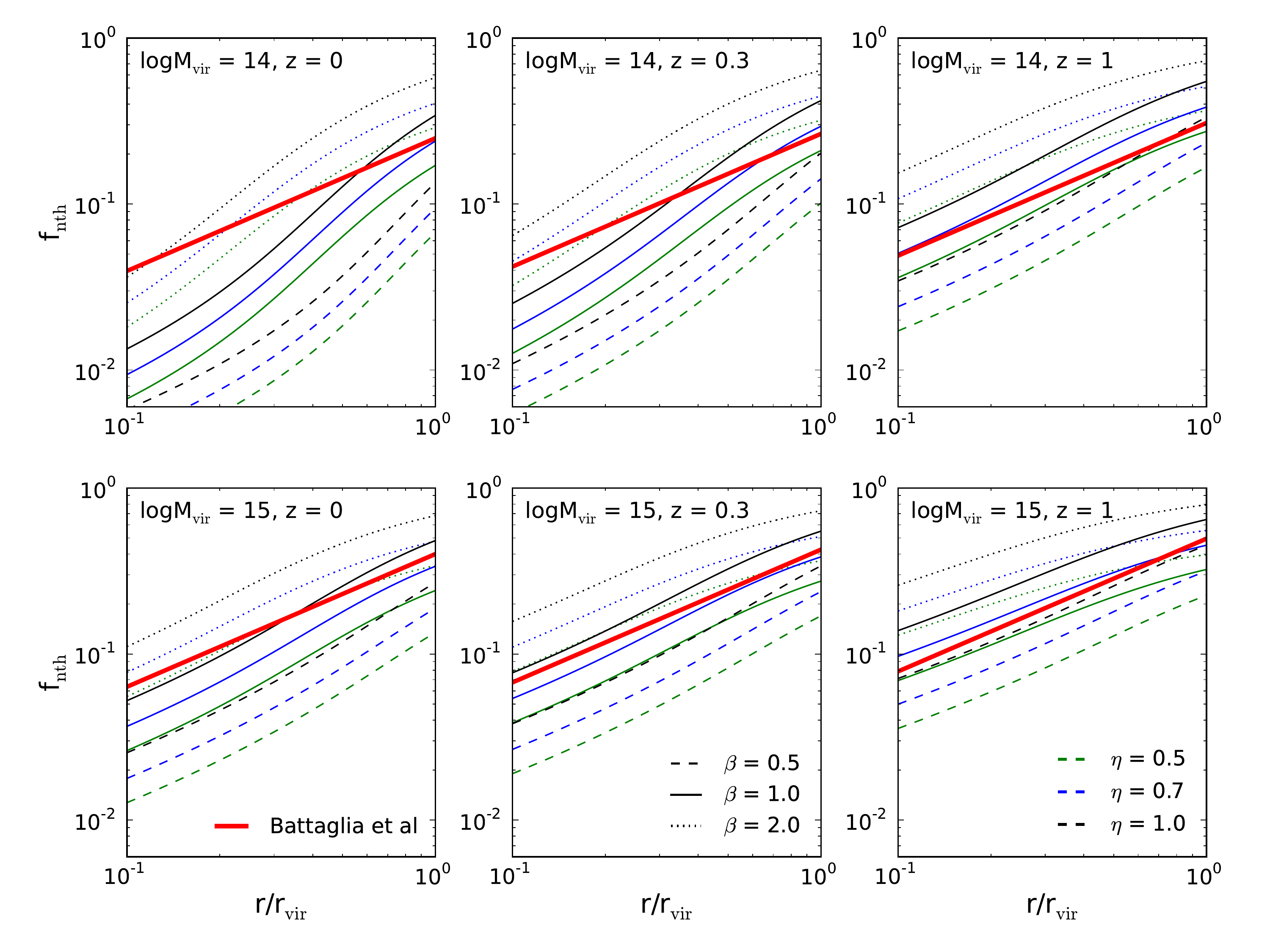}
\caption{Predicted non-thermal fraction, $f_{\rm nth}$, of representative galaxy
clusters with the mean mass growth history. The upper and
 lower panels  show
 $f_{\rm nth}$ for clusters with $M_{\rm vir} = 10^{14}$
 and $10^{15}~h^{-1}~$M$_{\sun}$, respectively. 
The left, middle,  and right panels show
three different redshifts $z=0$, 0.3, and 1. 
In each panel, the dashed, solid, and dotted lines show
 $\beta=0.5$, 1, and 2, respectively, while the green, blue, and black
 lines show $\eta=0.5$, 0.7, and 1, respectively.
The non-thermal fraction increases linearly with 
 $\eta$, and roughly linearly with $\beta$.
The fitting formula given by \citet{bat12} is shown as the
 thick solid  line.  
}
\label{fig:fnth}
\end{figure*}

Fig.\;\ref{fig:fnth} shows the value of the non-thermal fraction,
$f_{\rm nth}$, for galaxy clusters of $M_{\rm vir}=10^{14}~h^{-1}~M_{\sun}$ and $10^{15}~h^{-1}~M_{\sun}$ at three redshifts
$z=0$, 0.3 and 1. 
The radial dependence of the predicted $f_{\rm nth}$ is primarily a reflection
of the radial dependence of the turbulence dissipation time-scale, $t_{\rm d}$.
The higher density, hence the smaller $t_{\rm d}$,
in the inner region of a cluster makes it harder for turbulence to grow; thus, 
the predicted $f_{\rm nth}$ increases with radii.

The same argument also explains the dependence of the predicted $f_{\rm
nth}$ on $\beta$. 
The larger $\beta$ is, the larger $t_{\rm d}$ is, and the larger
the predicted $f_{\rm nth}$ becomes. 
Fig.\;\ref{fig:fnth} shows that $f_{\rm nth}$ is almost proportional to
$\beta$, until   
when $f_{\rm nth}$ starts to saturate to its long-term upper limit set by the
turbulence injection efficiency, $\eta$. 
As shown in Sect.\;\ref{sec:analytical}, when the initial condition is no longer
important, $f_{\rm nth}\propto \eta$, which explains
the dependence on $\eta$ presented in
Fig.\;\ref{fig:fnth}.

The mass and redshift dependence of $f_{\rm nth}$ originate from 
different mass growth histories of clusters. To the first order, the increase
of $f_{\rm nth}$ with 
cluster masses and redshifts can be understood as a result of the higher
mass growth rates (hence smaller $t_{\rm growth}$) at high redshifts and
for more massive clusters.
The radial dependence of $f_{\rm nth}$ also depends on masses
and redshifts. As $f_{\rm nth}$ approaches its limiting
value, the radial dependence weakens.

\subsection{Comparison with hydrodynamical simulations}
\label{sec:simcomp}
Hydrodynamical simulations help us further constrain the range of  
the model parameters. In two independent sets of simulations
\citep{shaw10,bat12}, the non-thermal fraction is found to be $\gtrsim
50\%$ at the cluster outskirts.  
There is no evidence for saturation
to a certain value. From this, the turbulence injection efficiency is
constrained to be $0.5\le \eta \le 1$, in accordance with the
expectation from the high turbulence injection efficiency associated with the
low Mach numbers of the internal shocks.
Fig.\;\ref{fig:fnth} shows that $f_{\rm nth}$ increases roughly linearly with the product of $\eta$ and $\beta$, as predicted by equation~(\ref{eq:fnthlim}) for $t_{\rm growth} > t_{\rm d}$. Even when $\eta$ is set to its upper limit, $\beta=0.5$  
cannot give $f_{\rm nth}$ 
seen in simulations. Therefore $\beta > 0.5$ is expected.

Recent simulation studies \citep{shaw10,bat12} 
provide fitting formulae of $f_{\rm nth}$.  
\citet{shaw10} use $16$ clusters in adaptive mesh refinement (AMR) simulations
\citet{nagai07}, whereas \citet{bat12} use a larger sample of galaxy clusters
from TreePM-SPH simulations. 
While their results agree at $z=0$, \citet{bat12} find 
stronger redshift evolution of the non-thermal pressure
\citep[see fig.\;3 of][and their text]{bat12}. 

We find that $f_{\rm
nth}$ with $\beta \approx 1$ and $\eta \approx 1$ agree with
the  fitting formula of \citet{bat12} (thick red solid lines in
Fig.\;\ref{fig:fnth}). 
Given that the fitting formula assumes factorizable dependence on the
radius, redshift, and mass, and that the typical mass\footnote{A typical mass of
$M_{200}=3\times 10^{14}~$M$_{\sun}$ is chosen by \citet{bat12}.} is chosen to
lie between those shown in Fig.\;\ref{fig:fnth}, we conclude that
the predicted radial slope also matches well with the simulations.
This strongly supports our explanation for the radial dependence of
$f_{\rm nth}$; namely, it is determined by the dependence of the
turbulence dissipation time on radii, i.e., $t_d$ increases rapidly with
radii from the inner regions to the outskirts.  

Dependence of $f_{\rm nth}$ on redshifts and masses 
found in numerical simulations is also 
consistent with our analytical predictions. 
The redshift dependence of $f_{\rm nth}$ predicted by our
model appears to be stronger than the fitting formula shows. However,
note that the simulation of 
\citet{bat12} actually shows stronger redshift dependence
than that is captured by the fitting formula. 

\subsection{Comparison with previous analytical work}

\citet{cav11} assume that the amplitude
of the turbulence pressure is determined by the residual kinetic energy of
the gas inflow after crossing the virial accretion shock, based on a smooth
accretion model of \citet{tozzi01} and \citet{voit03}.
The weakening of the accretion shocks at lower redshift due to the subsided
accretion rate suggests a lower shock heating efficiency, hence a higher
turbulence injection efficiency. Therefore, in their picture, the turbulence
pressure is expected to become important only at late cosmic epochs, $z<0.3$.
This is in complete contrast to our results, which show that the subsided
accretion rate at lower redshifts leads to a {\it decreasing} non-thermal
fraction. The increase of the non-thermal fraction with redshift seen in
the simulations of \citet{bat12} agrees with our results. 

Our model is highly predictive. We can calculate 
the radius, mass, and redshift dependence of the non-thermal pressure,
once two physical dimensionless parameters, $\beta$ and $\eta$, are
specified. \citet{cav11} parametrize the non-thermal fraction as $P_{\rm
nth}/P_{\rm th}=\delta_R\exp[-(r_{\rm vir}-r)^2/\ell^2]$, treating
$\delta_R$ and $\ell$ as free parameters. While these parameters must
at least 
depend on masses and redshifts, no prescriptions for computing these
parameters from their turbulence injection scenario are
given. \cite{bode/etal:2012} use the fitting formulae of \cite{shaw10}
for the radial dependence of the non-thermal pressure, and set the
amplitude according to dynamical state of clusters; thus, their approach
is more phenomenological than ours.

More specific
tests of our model using simulations and observations will be discussed
in Sect.\;\ref{sec:test}.

\subsection{Thermal pressure}  

Combining our model of the non-thermal fraction
and the model of the total pressure, 
we can calculate the observable quantity: the thermal pressure profile. 

 X-ray observations of a 
sample of nearby clusters \citep{arnaud10} have
shown that the 
intracluster thermal pressure profile follows an
approximately universal shape out to a radius of $r_{500}$ within which
the mean mass density is 500 times the critical density. 
This `universal thermal pressure profile' agrees  roughly with the
profiles seen in numerical simulations \citep{bor04,nagai07b,piff08}
after considering a $15\%$ mass bias between the two.

Recent observations of the SZ effect by the Planck satellite
 provide a direct measurement of the stacked thermal pressure
profile of intracluster gas out to large radii \citep{planck13a}. They find a
good agreement with the X-ray-derived profiles in the inner regions,
 while the Planck-derived profiles are shallower than the
 X-ray-derived ones at the outskirts.

\begin{figure}
\centering
\includegraphics[width=0.45\textwidth]{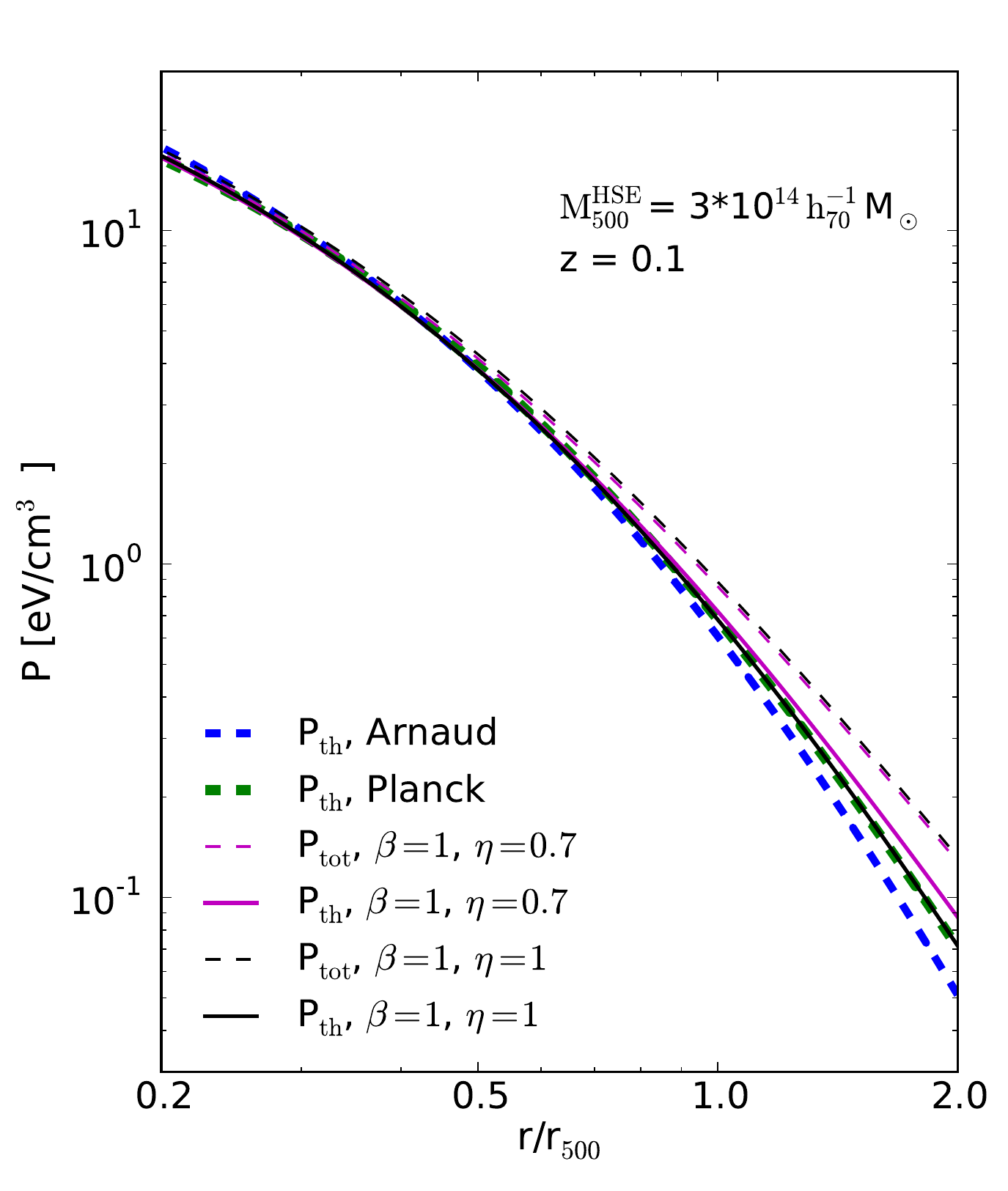}
\caption{Predicted total (thin dashed) and thermal (solid) pressure profiles
 compared to observations. The blue (lower) and green (upper) thick dashed
 lines show the profiles derived
 from X-ray  \citep{arnaud10} and SZ \citep{planck13a} observations,
 respectively. Note that the total pressure depends very weakly on $\beta$ and
 $\eta$, because the derived true cluster masses differ for different
 non-thermal pressure fraction.}
\label{fig:Ptot}
\end{figure}

Fig.\;\ref{fig:Ptot} shows a comparison of the total and thermal pressure
profiles predicted by our model and the observations.\footnote{The electron
pressure profiles derived from the observations
are converted to the pressure
contributed by all particles with the number density conversion factor
of $n_{\rm e}/n = 0.52$ for fully ionized gas with the
hydrogen abundance of $76\%$ and the helium abundance of 24\%.} 
The predicted total pressure is significantly higher than the thermal
pressure derived from the observations at the cluster outskirts.
The difference between them increases
towards larger radii, suggesting larger non-thermal pressure support
there. The thermal 
pressure profiles we compute from our model, with preferred values of the
parameters of $(\beta,\eta)=(1,1)$ or $(1,0.7)$, agree
remarkably well with the observations. Our
model has passed an important observational test.

\section{Hydrostatic mass bias}
\label{sec:bias}
 
\begin{figure}
\centering
\includegraphics[width=0.45\textwidth]{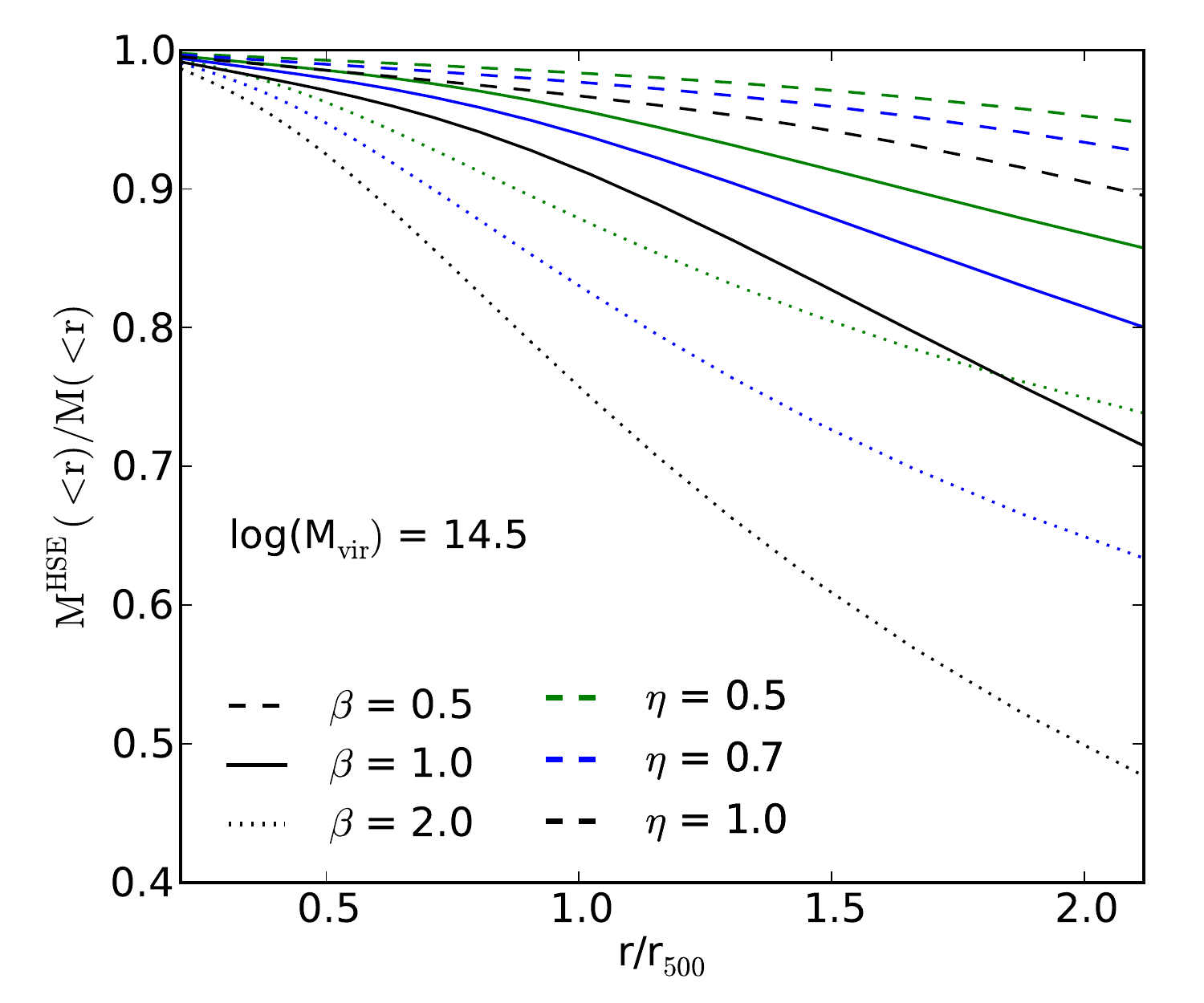}
\caption{Predicted hydrostatic mass bias for a galaxy cluster at $z=0$
 with the mass of $10^{14.5}~h^{-1}~M_{\sun}$ as a function of
 $r/r_{500}$. The dashed, solid, and dotted lines show
 $\beta=0.5$, 1, and 2, respectively, while the green, blue, and black
 lines show $\eta=0.5$, 0.7, and 1, respectively.}
\label{fig:bias1}
\end{figure}

\begin{figure}
\centering
\includegraphics[width=0.45\textwidth]{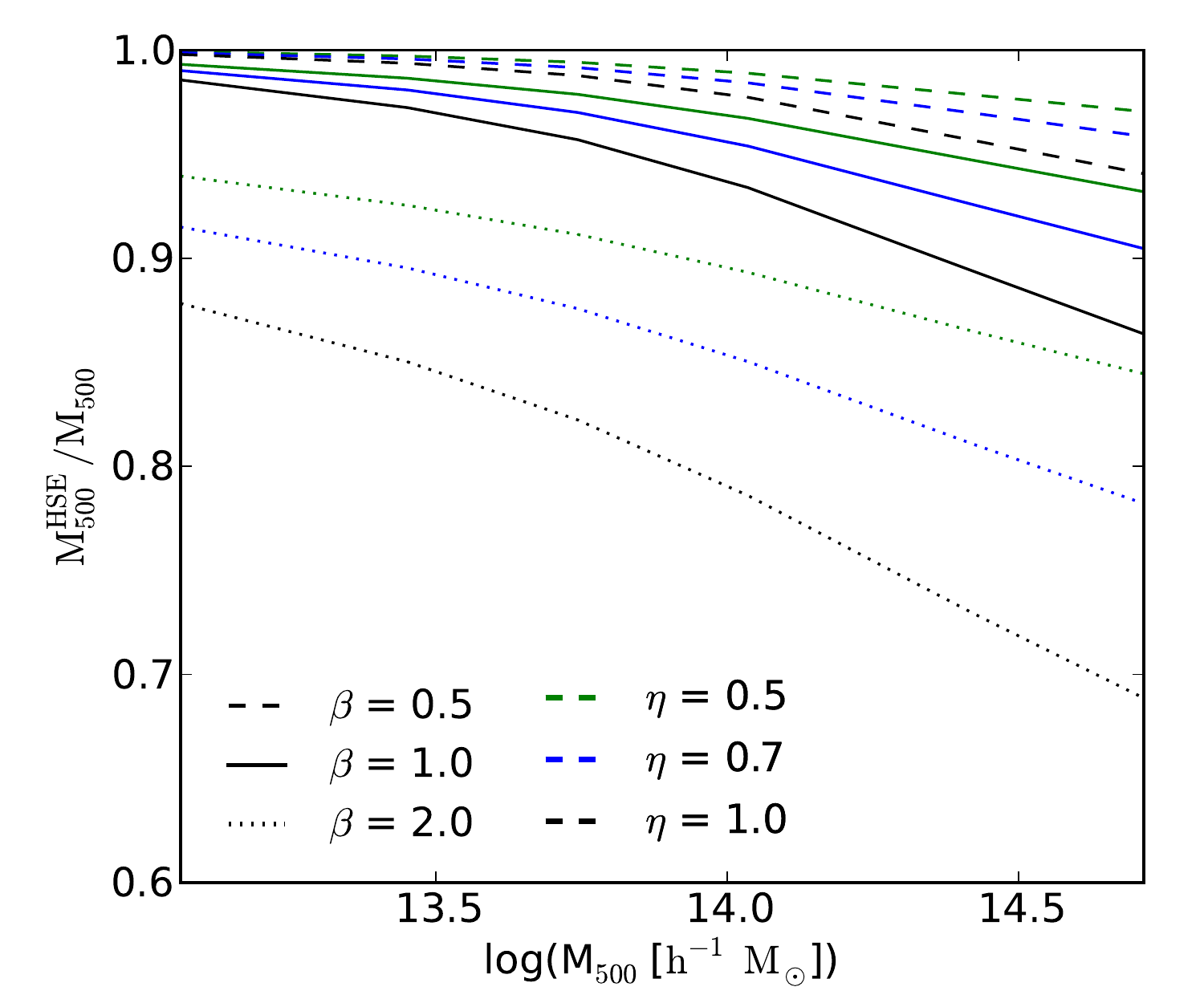}
\caption{Similar to Fig.~\ref{fig:bias1}, but for the predicted hydrostatic
 mass bias for the mass enclosed within  $r_{500}$ as a function of the
 true $M_{500}$. 
} 
\label{fig:bias2}
\end{figure}

It is common practice to use X-ray data of galaxy clusters to
infer their masses assuming hydrostatic equilibrium between the thermal
pressure and gravity. The presence of non-thermal pressure inevitably
biases such mass estimates. Our model allows us to calculate this
`hydrostatic mass bias.' Specifically, we calculate the mass enclosed
within a given radius $r$ inferred from hydrostatic equilibrium, $M^{\rm
HSE}(<r)$, and compare it to the true mass, $M(<r)$. By definition,
$M^{\rm HSE}(<r)\le M(<r)$ at all radii.

Fig.~\ref{fig:bias1} shows the ratios of $M^{\rm HSE}$ and the true
mass as a function of $r/r_{500}$ for various combinations of $\beta$
and $\eta$. The true virial mass of the cluster is
$10^{14.5}~h^{-1}~M_{\sun}$. The hydrostatic mass bias increases towards
large radii due to the increased non-thermal pressure support. 
We typically find 5-10\% and 20-30\% mass biases for $r\approx r_{500}$
and $2r_{500}$, respectively, for plausible parameters,
$(\beta,\eta)=(1,1)$ and $(1,0.7)$.

Fig.~\ref{fig:bias2} shows the ratios of $M^{\rm HSE}$ and the true
mass enclosed within $r=r_{500}$ as a function of the true
$M_{500}\equiv M(<r_{500})$ for various combinations of $\beta$
and $\eta$. The hydrostatic mass bias increases towards large masses due
to higher mass accretion rates. We typically find 10\% bias for rich
clusters for $(\beta,\eta)=(1,1)$ and $(1,0.7)$.

\section{Other sources of  deviation from hydrostatic equilibrium}
\label{sec:dis}
In our model  
the source of intracluster turbulence is the
growth of clusters via mergers and accretion. Another commonly
recognized source is the active galactic nuclei (AGNs) hosted by central
galaxies in  clusters, which can inject additional turbulence into the
cluster cores.  
This additional source  can be taken naturally into account by adding a
radius-dependent source term to the r.h.s. of
equation~(\ref{eq:sigkin}). In this paper we neglect this, and focus on
the outskirts of galaxy clusters.

The parameters $\beta$ and $\eta$ are taken as constants in our model. In
principle, they may vary with radii, cluster masses, and redshifts. For
example, $\eta$ is in general a weak function of radii, as some
simulations indicate that the Mach number of internal shocks varies
weakly with radii.
Once the forms of these dependencies are known, it is straightforward
to incorporate them into our model. Neglecting a possible dependence of
$\beta$ and $\eta$ on radii, masses, and redshifts would introduce an
error in our model. 
However, the $t_{\rm d}$-induced radial dependence and the $t_{\rm
growth}$-induced cluster mass and redshift dependence are much stronger,
and  additional dependencies are expected to be negligible compared to them.

So far, we have assumed the turbulence pressure to be isotropic. Whereas
an isotropic state usually is a higher entropy state to which a system
generally evolves, physical conditions, for instance the density
stratification in the intracluster medium, can trigger an anisotropic velocity
field. The degree of anisotropy induced by the density stratification can be
estimated 
by comparing the buoyancy time-scale and the turbulence eddy turnover time
scale. The corresponding dimensionless number, the Froude number, is found to be
close to, or greater than, $1$ in a typical cluster environment
\citep{gas13,gas14}, implying that the anisotropy is weak. \citet{vazza12} use
hydrodynamical simulations to  demonstrate that anisotropy of the turbulence velocity
field is not particularly strong in the virial region of relaxed clusters.
Anisotropic magnetic field, if exists, can also induce anisotropic turbulence
pressure. This is beyond the scope of this paper. 
In any case, in the presence of anisotropy,
the non-thermal fraction given in this paper should be interpreted as
the ratio of the radial components of $\sigma^2_{\rm nth}$ and
$\sigma^2_{\rm tot}$, whose evolution is described by equation
(\ref{eq:sigkin}).

Other sources of pressure support in the intracluster gas include 
magnetic fields and cosmic rays.  Non-thermal
cluster phenomena such as radio haloes, radio relics, and non-thermal X-ray
emission \citep[e.g.,][]{nev04,bru08,million09,fer12,kale13} have been observed in some
nearby galaxy clusters, suggesting the existence of magnetic fields and cosmic
ray electrons in those clusters. 
The inferred amplitude of magnetic fields is on
the order of $\mu$G, but its origin, universality and the non-thermal pressure
it gives are still under debate \citep[see, e.g.,][]{dolag00, car02, iapichino12}.  
If, as commonly anticipated, magnetic fields in clusters are amplified by
turbulence dynamo \citep{ensslin06,kang07}, then its saturation time-scale is set by the
eddy turnover time-scale, and its energy dissipates as turbulence energy
dissipates. The non-thermal pressure it contributes to can then be modelled
conveniently by adding another source term to the r.h.s. of
equation~(\ref{eq:sigkin}), and its amplitude is limited by that of the
turbulence pressure.
Cosmic rays are expected to be generated in accretion and merger shocks
with high Mach numbers. 
However, efforts to detect signatures of cosmic ray ions in galaxy
clusters have so far resulted in null results, which limits the cosmic ray
pressure contribution to a few percent of the thermal pressure in several 
clusters \citep[e.g.,][]{ack10,ack13}.

Recently, some numerical studies \citep{lau09,lau13,suto13,nelson13} show that
residual, collective accelerated motion in the gas at cluster
outskirts contributes to additional deviation from hydrostatic equilibrium.
This `acceleration bias' cannot be taken into account as a form of pressure;
however, it also originates from recent merger/accretion of the cluster, and has
a dissipation time-scale of the sound crossing time which is comparable to the
dynamical time  defined in equation~(\ref{eq:td}). Therefore one
naturally expects 
that the acceleration bias increases towards larger radii due to
significantly longer sound crossing times there. This may enable a treatment of
the acceleration bias using the same framework of our model by considering its
effect at each Eulerian radius separately.

\section{Further tests of our model}
\label{sec:test}
The fitting formula for $f_{\rm nth}$ proposed by \citet{bat12}, to which
we compared our results in Sect.\;\ref{sec:simcomp}, is not optimal in
describing the simulation results since it assumes a factorizable form
in terms of radius, cluster mass, and redshift.
In general, these dependencies are coupled to each other, as found in both
simulations and our model. Since our model predicts how these 
dependencies couple, e.g., radial dependence of $f_{\rm nth}$ is stronger for low
redshift and less massive clusters, one can test the existence of the predicted
coupling in numerical simulations. 

We have compared our model with simulations and observations using the
average properties of cluster samples. With the help of hydrodynamical
simulations with high time resolution, the evolution of profiles of {\it
individual} galaxy clusters along their mass growth histories can be
measured. This would enable a far more detailed comparison between the
predicted non-thermal pressure and simulations of individual clusters.

Although comparison with observations is limited to cluster samples, here too,
exists a large room for improvements. For instance, the redshift
evolution and mass dependence of the HSE mass bias could be studied and
tested against the predicted trend. 

The biggest uncertainty in our non-thermal fraction model so far is the
uncertainty in two physical parameters: the turbulence injection efficiency
$\eta$, and especially $\beta$, the ratio between
the turbulence dissipation time and the local dynamical time. We have
constrained them to be $0.6<\eta\le 1$ and $0.5<\beta<2$ from comparison to
simulation results of non-thermal fraction and its mass dependence. We find that
 $\eta\beta\approx 1$ is preferred from the existing observations.
Numerical experiments of, e.g.,
turbulence generation by a single minor merger event and turbulence dissipation
in an isolated cluster, can help us further pin down their values, as well
as prove the basis of our theoretical model.

\section{Conclusion}
\label{sec:con}

We have developed an analytical model for non-thermal pressure of
intracluster gas. The key part of this model is a model of the
non-thermal fraction, $f_{\rm 
nth}$, which is based on a description of the evolution of turbulence energy in
the intracluster medium in the form of a first-order differential equation.

Three time-scales are  
responsible for  
the non-thermal fraction.
In particular, 
the ratio of $t_{\rm growth}$ and $t_{\rm d}$ determines the limiting value of
the non-thermal fraction, while the ratio of $t_{\rm obs}-t_{\rm i}$ and
min($t_{\rm growth},t_{\rm d}$) determines how much the value of $f_{\rm
nth}$ at $t_{\rm obs}$ has replaced its initial value and
approached its limiting value.

The other parts of our non-thermal pressure model include a model of the total
pressure, which is based on 
a re-interpretation of the self-similar model of gas pressure profile
presented in \citet{kom01}, and a model of the average mass growth
history of clusters presented in \cite{zhao09}.

The non-thermal fraction predicted by our model lies in the range seen in
numerical simulations. 
Generic trends of the non-thermal fraction, e.g., the
increase with radius, cluster mass and redshift,  
seen in numerical simulations are successfully  
reproduced by our analytical model. Our analytical model
gives physical insights into the cause of the generic trends.
For example, the increasing non-thermal
fraction with radii is explained by the increasing
turbulence dissipation time-scale,  
and the increase at higher redshift is mainly due to the higher cluster
growth rate.

Combining our model of the non-thermal fraction and the self-similar model of
the total pressure profile, we obtain thermal pressure profiles which match well
with the existing simulations and observations. We also
calculate the hydrostatic mass bias, typically finding 10\% mass bias
for $M_{500}$ of rich clusters.

Our model provides physical understandings as well as detailed predictions of
the non-thermal pressure support in galaxy clusters. Once validated by hydrodynamical
simulations of individual clusters and observations of different cluster samples
(in addition to validation presented in this paper), it promises more accurate
cluster mass estimations using HSE.

\section*{Acknowledgements}
We thank Klaus Dolag for helpful discussions, as well as giving us
 access to his simulation data. We are also grateful to many colleagues
 at MPA, in particular Eugene Churazov, Rishi Khatri, Friedrich Meyer
 and Simon White for related discussions. 
We thank Nick Battaglia, Hy Trac, and Peng Oh for comments on the draft.

\bibliographystyle{mn2e}

\appendix

\section[]{Compute the non-thermal pressure}
\label{app:procedure}

\begin{figure*}
\centering
\includegraphics[width=0.8\textwidth]{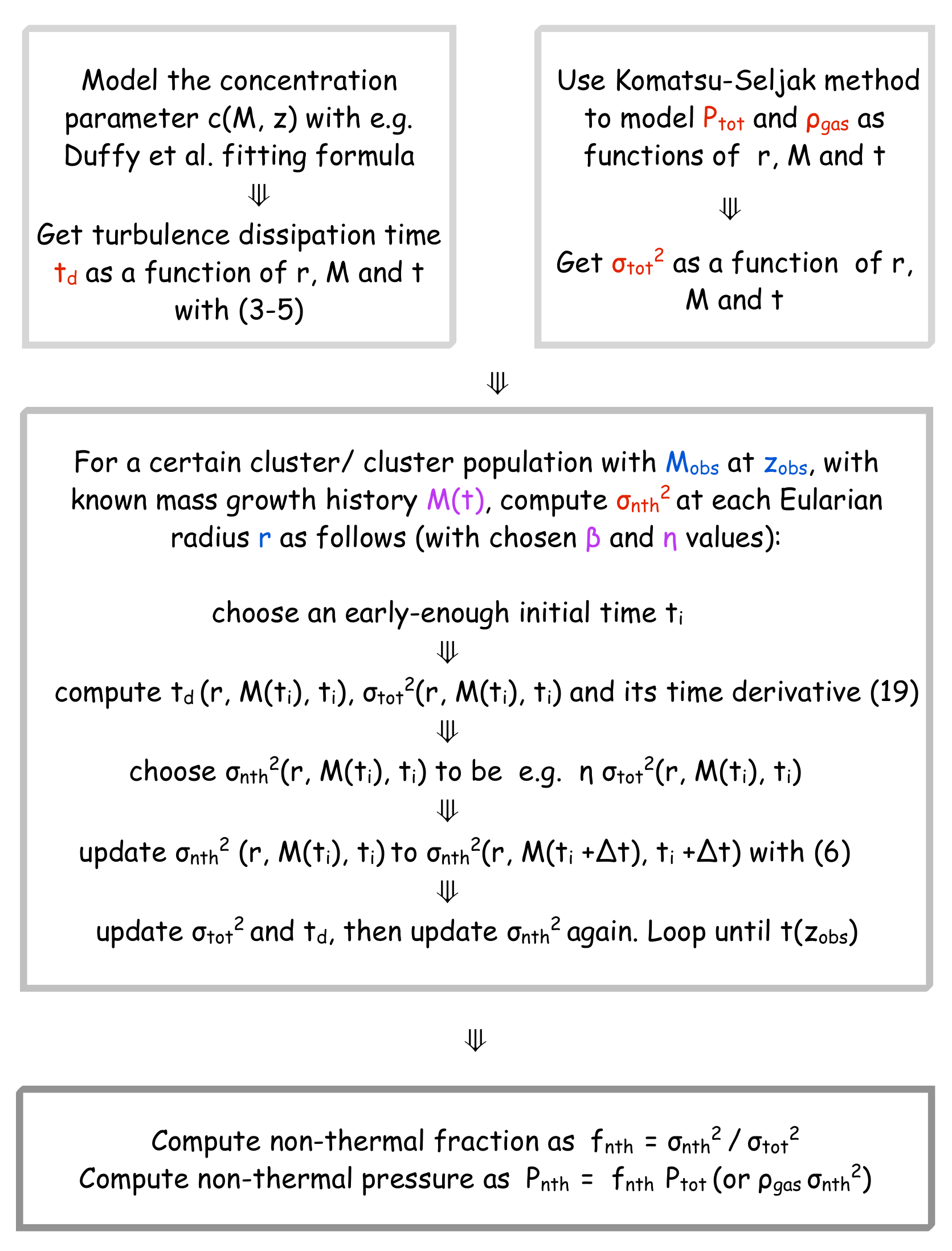}
\label{fig:flowchart}
\caption{Steps necessary for computing the non-thermal
 pressure. Important intermediate 
quantities are marked in red. The mass growth history, $M(t)$, and the two model
parameters, $\beta$ and $\eta$, which are marked in magenta, need to be
 chosen in advance. The final outputs ($f_{\rm nth}$, $P_{\rm nth}$) are given as
functions of the Eulerian radius $r$, and depend on the final mass and redshift
of the cluster, which are marked in blue. }
\end{figure*}

Fig.\;\ref{fig:flowchart} is a flowchart of the steps necessary for
computing the non-thermal pressure. 
The key part of our non-thermal pressure model is the first-order differential
equation given in equation~(\ref{eq:sigkin}). The steps of solving this equation
are shown in the box in the middle of Fig.\;\ref{fig:flowchart}. Two
inputs, namely 
the turbulence dissipation time $t_{\rm d}$ and the total velocity dispersion
$\sigma_{\rm tot}$ as functions of radius, cluster mass and redshift, are
required before solving the differential equation. 

The procedures we have adopted in this paper for modelling $t_{\rm d}$ and 
$\sigma_{\rm tot}$ are shown as the two separate boxes at the top of
Fig.\;\ref{fig:flowchart}. The turbulence dissipation time, $t_{\rm d}$, is
proportional to the dynamical time-scale which is determined by the depth of the
gravitational potential well. We model the gravitational potential as a function
of radius, cluster mass and redshift with that of an NFW profile. The total
velocity dispersion, $\sigma_{\rm tot}$, is calculated from the
Komatsu-Seljak profile\footnote{Fortran routines are
provided at
\url{http://www.mpa-garching.mpg.de/~komatsu/CRL/clusters/komatsuseljakprofile}}
\citep{kom01}, 
which is re-interpreted as a model for the total pressure and the gas
density profiles. 
Other models of $t_{\rm d}$ and/or $\sigma_{\rm tot}$ may be used instead
of these models.

Solving equation~(\ref{eq:sigkin}) also requires a known mass growth history
$M(t)$ of the target galaxy cluster or cluster sample. In this paper we
consider a representative sample of galaxy clusters; namely, those with
the average mass growth history of all clusters of a given mass and
redshift. For the average mass growth history we take the model of
\citet{zhao09}.\footnote{Fortran routines are provided at \url{http://www.shao.ac.cn/dhzhao/mandc.html}}

The final outputs, $f_{\rm nth}$ and $P_{\rm nth}$, can be computed
directly from the solution of equation~(\ref{eq:sigkin}), as shown in
the box at the bottom of Fig.\;\ref{fig:flowchart}.

\label{lastpage}
\end{document}